\documentclass[11pt,english]{article}
\usepackage[T1]{fontenc}
\usepackage{a4wide}
\usepackage{babel}
\usepackage{graphics}

\makeatletter

\newcommand{\LyX}{L\kern-.1667em\lower.25em\hbox{Y}\kern-.125emX\spacefactor1000}

\makeatother

\begin{document}

{\centering \textbf{\huge Color superconductivity, \( Z_{N} \) flux tubes and
monopole confinement in deformed \( \mathcal{N}=2^{*} \) super Yang-Mills theories}\huge \par}

\begin{center}

\vspace{1cm}

{\large {\bf Marco A. C. Kneipp}}\footnote{Email address: {kneipp@cbpf.br}} 

\vspace{0.3cm}

\end{center}

\begin{center}{\em Universidade do Estado do Rio de Janeiro(UERJ),\\

Dept. de Física Teórica,\\

Rua São Francisco Xavier, 524\\

20550-013 Rio de Janeiro, Brazil.\\ 
\medskip{}

Centro Brasileiro de Pesquisas F\' \i sicas (CBPF), \\ 

Coordenação de Teoria de Campos e Partículas (CCP), \\ Rua Dr. Xavier Sigaud,
150 \\

22290-180 Rio de Janeiro, Brazil. \\ }

\end{center}

\begin{abstract}
We study the \( Z_{N} \) flux tubes and monopole confinement in deformed \( \mathcal{N}=2^{*} \)
super Yang-Mills theories. In order to do that we consider an \( \mathcal{N}=4 \)
super Yang-Mills theory with an arbitrary gauge group \( G \) and add some
\( \mathcal{N}=2 \), \( \mathcal{N}=1 \), and \( \mathcal{N}=0 \) deformation
terms. We analyze some possible vacuum solutions and phases of the theory, depending
on the deformation terms which are added. In the Coulomb phase for the \( \mathcal{N}=2^{*} \)
theory, \( G \) is broken to \( U(1)^{r} \) and the theory has monopole solutions.
Then, by adding some deformation terms, the theory passes to the Higgs or color
superconducting phase, in which \( G \) is broken to its center \( C_{G} \).
In this phase we construct the \( Z_{N} \) flux tubes ansatz and obtain the
BPS string tension. We show that the monopole magnetic fluxes are linear integer
combinations of the string fluxes and therefore the monopoles can become confined.
Then, we obtain a bound for the threshold length of the string breaking. We
also show the possible formation of a confining system with three different
monopoles for the \( SU(3) \) gauge group. Finally we show that the BPS string
tensions of the theory satisfy the Casimir scaling law.

\vfill PACS numbers: 11.27.+d, 11.15.-q, 02.20.-a 
\end{abstract}
\newpage

\section{Introduction}

It has long been believed that particle confinement at the strong coupling regime
should be a phenomenon dual to monopole confinement in a (color)superconductor
in weak coupling. Therefore, the study of monopole confinement in weak coupling
may shed some light on particle confinement and other phenomena in the dual
confined theory. Since dualities are better understood for supersymmetric theories
and, in particular, for finite ones, it is interesting to consider the monopole
confinement in these theories or deformations of them. Therefore in the present
work we study monopole confinement in an \( \mathcal{N}=2^{*} \) super Yang-Mills
theory with the addition of some \( \mathcal{N}=1 \) or \( \mathcal{N}=0 \)
deformation terms.

Since the papers of Seiberg and Witten \cite{sw1, sw2}, quite a lot work \cite{douglas shenker}-\cite{tong}
has been done analyzing different aspects of confinement in supersymmetric theories.
Usually, one starts with a microscopic \( \mathcal{N}=2 \) \( SU(N) \) super
Yang-Mills (SYM) theory (with some possible matter fields) and then obtains
an effective \( \mathcal{N}=2 \) \( U(1)^{N-1} \) SYM theory with an \( \mathcal{N}=1 \)
deformation term. In this theory, each \( U(1) \) factor is broken to \( Z \),
resulting in \( (N-1) \) infinite towers of Nielsen-Olesen flux tubes or strings,
which gives rise to confinement of Dirac monopoles. However, as was pointed
out in \cite{strassler}, a similar phenomenon is not expected to happen to
quark confinement in QCD. It is then believed that only some of these strings
might be stable, which could correspond to \( Z_{N} \) strings in the microscopic
theory.

On the other hand, in \cite{kb, k} solitonic monopoles and \( Z_{k} \) strings
were obtained directly as solutions of the same theory with two gauge symmetry
breakings. In order to do that, we considered \( \mathcal{N}=2 \) super Yang-Mills
theories with an arbitrary simple gauge group \( G \), a massive hypermultiplet,
and an \( \mathcal{N}=0 \) deformation mass term. This hypermultiplet was considered
to be in the symmetric part of the tensor product of \( k \) fundamental representations,
with \( k\geq 2 \). We considered this theory in weak coupling and showed the
existence of vacuum solutions which produce the symmetry breaking
\[
G\, \rightarrow \, G_{S}\equiv [G_{0}\times U(1)]/Z_{l}\, \rightarrow \, G_{\phi }\equiv [G_{0}\times Z_{kl}]/Z_{l,}\]
where \( G_{0} \) is a subgroup of \( G \) and \( Z_{l} \) is a common discrete
subgroup of \( G_{0} \) and \( U(1) \) as explained in \cite{kb, k}. The
first symmetry breaking happens when the \( \mathcal{N}=0 \) deformation mass
parameter \( m \) vanishes. Then, the theory has solitonic monopoles which
should fill representations of \( G^{\textrm{v}}_{0} \), the dual group of
\( G_{0} \). The second symmetry breaking happens when \( m>0 \). Since in
this phase \( \Pi _{1}(G/G_{\phi })=Z_{k} \), there exist \( Z_{k} \) strings
or flux tubes. Moreover, since the \( U(1) \) factor is broken, the monopole
magnetic lines in this \( U(1) \) direction can no longer spread radially over
space. However, since the monopole magnetic flux is an integer multiple of the
``fundamental'' \( Z_{k} \) string flux, these lines can form \( Z_{k} \)
strings and monopoles become confined.  

It is interesting to note that, when \( k=2 \), the complex scalar \( \phi  \)
which produces the second symmetry breaking that allows the existence of \( Z_{k} \)
strings is in the same representation as that of a diquark condensate. One then
could think of \( \phi  \) as being itself this diquark condensate, and therefore
we would have a situation quite similar to the one in an ordinary superconductor,
described by the Abelian Higgs theory with the scalar being a Cooper pair. In
addition, if the gauge group is \( SU(N) \), the scalar in the adjoint representation
of the vector supermultiplet could also be thought to be a quark-antiquark condensate.
These two kinds of condensates are indeed expected to exist in the color superconducting
phase of (dense) QCD at the weak coupling \cite{colorsupercondunting, colorsuper2}.
The effective theory describing these condensates is not well known. It should
be a \( SU(3) \) Yang-Mills-Higgs theory with some scalars in color sextet
and color octet representations. Therefore, one could think that the theory
used in \cite{kb, k} or in the present paper, when the gauge group is \( G=SU(3) \),
as been a toy model for an effective theory of these condensates. Then, one
can conclude that the effective theory for these condensates could have monopoles,
flux tubes, and monopole confinement, depending on the form of the potential. 

Although the monopoles in \cite{k} should fill representations of the non-Abelian
group \( G^{\textrm{v}}_{0} \), monopole confinement happened through flux
tubes in a \( U(1) \) direction inside the non-Abelian group \( G \). The
motivation of the present paper is to consider monopole confinement through
the formation of flux tubes due to breaking of the full non-Abelian group \( G \),
although in the present case (stable) monopoles are not expected to fill representations
of non-Abelian groups. In order to do that, we shall consider the bosonic part
of \( \mathcal{N}=4 \) SYM theory in the weak coupling regime and add some
\( \mathcal{N}=2 \), \( \mathcal{N}=1 \), or \( \mathcal{N}=0 \) deformation
mass terms. These SYM theories are usually denoted by \( \mathcal{N}=2^{*} \),
\( \mathcal{N}=1^{*} \), and \( \mathcal{N}=0^{*} \), respectively. In \cite{witten,  strassler}
was pointed out that the \( \mathcal{N}=1^{*} \) theory should have a weakly
coupled Higgs phase with magnetic flux tubes and this phase should be dual to
a strongly coupled confining phase in the dual theory. One of the aims of the
present paper is to analyze some properties of these magnetic flux tubes. In
section 2, we obtain the lower bound for the string tension and corresponding
Bogomol'nyi-Prasad-Sommerfield (BPS) string conditions for a Yang-Mills theory
with three complex scalars in the adjoint representation. Then, in section 3,
we analyze the possible vacuum solutions and corresponding gauge symmetry breaking
which happen depending on the mass deformation terms which are added to the
\( \mathcal{N}=4 \) super Yang-Mills theory. We show that in this theory there
are vacuum solutions which produce the spontaneous symmetry breaking,
\[
G\, \rightarrow \, U(1)^{r}\, \rightarrow \, C_{G},\]
where \( r \) is the rank of \( G \) and \( C_{G} \) its center. The first
symmetry breaking happens in the \( \mathcal{N}=4 \) and \emph{\( \mathcal{N}=2^{*} \)}
theories. Then, the second symmetry breaking happens when one adds to the \( \mathcal{N}=2^{*} \)
theory an \( \mathcal{N}=1 \) or an \( \mathcal{N}=0 \) deformation term (or
both). In section 4, we analyze the Coulomb or free-monopole phase which occurs
in the first symmetry breaking. In this phase there are BPS monopole solutions.
In section 5, we analyze the Higgs or color superconducting phase which occurs
when the second symmetry breaking happens. In this phase the monopoles chromomagnetic
lines cannot spread out radially over space. However, since in this phase 
\[
\Pi _{1}(G/C_{G})=C_{G},\]
 when \( C_{G} \) is nontrivial, these flux lines can form topologically nontrivial
\( Z_{N} \) strings. We then construct the \( Z_{N} \) string ansatz. Some
\( Z_{N} \) string solutions have been considered in \cite{znvortices} for
different \( SU(N) \) gauge theories. We show that the flux of the magnetic
monopoles can be expressed as an integer linear combination of the string fluxes.
Therefore, in the Higgs phase the monopole magnetic lines can form \( Z_{N} \)
strings and the monopole can become confined, as in \cite{k}. We then obtain
for the monopole-antimonopole system a bound for the threshold length for the
string-breaking. In section 6 we consider \( G=SU(N) \) and analyze how the
monopole magnetic flux could be considered to be formed by a set of a string
and an antistring in the fundamental representation. For the \( SU(3) \) gauge
group we show that, besides the monopoles-antimonopole system, the monopoles
with strings attached could form a confining system with three different monopoles.
In section 7, we show that the BPS string tensions satisfy the Casimir scaling
law.

\section{String BPS conditions}

Let us start with a Yang-Mills-Higgs theory with three complex scalars \( \phi _{s},\, s=1,\, 2,\, 3 \),
in the adjoint representation of an arbitrary gauge group \( G \) which we
shall consider to be simple, connected, and simply connected. Let \( \phi _{s}=M_{s}+iN_{s} \)
where \( M_{s} \) and \( N_{s} \) are real scalars and pseudo-scalars, respectively.
Let us consider the Lagrangian
\[
L=-\frac{1}{4}G_{a\mu \nu }G_{a}^{\mu \nu }+\frac{1}{2}\left( D_{\mu }\phi _{s}^{*}\right) _{a}\left( D^{\mu }\phi _{s}\right) _{a}-V\]
where \( V \) is for the moment an arbitrary positive potential. In \cite{kb},
a theory was considered with a complex scalar in the adjoint and another complex
scalar initially in an arbitrary representation, and the non-Abelian string
BPS conditions for an arbitrary gauge group were obtained. Let us repeat this
procedure for the case with three scalars in the adjoint. Let \( D_{\mu }=\partial _{\mu }+ieW_{\mu } \),
\( D_{\pm }=D_{1}\pm iD_{2} \), and \( B_{i}=-\varepsilon _{ijk}G_{jk}/2 \)
is the non-Abelian magnetic field. Let us consider a static configuration with
cylindrical symmetry and not depending on \( x_{3} \). Then, generalizing the
Bogomol'nyi procedure \cite{bogo}, we obtain that the string tension \( T \)
must satisfy 
\begin{eqnarray*}
T & = & \int d^{2}x\left\{ \frac{1}{2}\left[ \left( E_{ia}\right) ^{2}+\left( B_{ia}\right) ^{2}+\left| \left( D_{\mu }\phi _{s}\right) _{a}\right| ^{2}\right] +V\right\} \\
 & \geq  & \int d^{2}x\left\{ \frac{1}{2}\left[ \left| \left( D_{\mp }\phi _{s}\right) _{a}\right| ^{2}+\left( B_{3a}\right) ^{2}\pm e\left( \phi ^{*}_{sb}if_{abc}\phi _{sc}\right) B_{3a}\right] +V\right\} \\
 & \geq  & \int d^{2}x\left\{ \frac{1}{2}\left( B_{3a}\right) ^{2}\pm d_{a}B_{3a}\pm X_{a}B_{3a}+V\right\} \\
 & = & \int d^{2}x\left\{ \frac{1}{2}\left[ B_{3a}\pm d_{a}\right] ^{2}\pm X_{a}B_{3a}-\frac{1}{2}\left( d_{a}\right) ^{2}+V\right\} ,
\end{eqnarray*}
where 
\begin{equation}
\label{2.1}
d_{a}\equiv \frac{e}{2}\left( \phi ^{*}_{sb}if_{abc}\phi _{sc}\right) -X_{a},
\end{equation}
 and the quantity \( X_{a} \) is a real scalar which transforms in the adjoint
representation. We could consider that 
\[
X_{a}=\frac{e}{2}\left[ m_{N_{s}}\textrm{Im}\left( \phi _{sa}\right) +m_{M_{s}}\textrm{Re}\left( \phi _{sa}\right) +c\delta _{a,0}\right] ,\]
where \( m_{N_{s}} \), \( m_{M_{s}} \) and \( c \) are real mass parameters
and the last term could only exist if \( G \) contains a \( U(1) \) factor
generated by \( T_{0} \) (and therefore \( G \) would not be simple). If
\begin{equation}
\label{2.2}
V\geq \frac{1}{2}\left( d_{a}\right) ^{2}\, ,
\end{equation}
it follows that 
\begin{equation}
\label{2.3}
T\geq \pm \int d^{2}xX_{a}B_{3a}\, .
\end{equation}
Since \( T\geq 0 \), we take the upper (lower) sign if the above integral is
positive (negative). The equality happens when
\begin{eqnarray}
B_{3a} & = & \mp d_{a,}\label{2.4a} \\
D_{\mp }\phi _{s} & = & 0,\label{2.4b} \\
V-\frac{1}{2}\left( d_{a}\right) ^{2} & = & 0,\label{2.4c} \\
E_{ia} & = & B_{1a}=B_{2a}=D_{0}\phi _{s}=D_{3}\phi _{s}=0,\label{2.4d} 
\end{eqnarray}
and we recover the non-Abelian string BPS conditions  in \cite{kb} for the
particular case in which all scalars are in the adjoint. As in \cite{kb}, for
simplicity we shall consider that \( m_{M_{3}} \) could be the only nonvanishing
mass parameter in \( X_{a} \) and we shall rename it \( m \). Note that if
we had chosen to set that only \( m_{N_{3}}\neq 0 \), then \( X_{a} \) would
allow a nonvanishing pseudoscalar vacuum solution which would result in the
magnetic charge of the monopole and the flux of the string being Lorentz scalars
and not pseudoscalars as usual. Moreover, we shall consider \( G \) to be simple
since we are interested in string solutions associated with the breaking of
a non-Abelian group and not due to the breaking of \( U(1) \) factors. Hence
\( X_{a} \) does not have the term \( c\delta _{a,0} \). Threfore, we shall
consider 
\[
X_{a}=\frac{em}{2}\textrm{Im}\left( \phi _{3a}\right) .\]

We shall consider the potential
\begin{equation}
\label{2.5}
V=\frac{1}{2}\left[ \left( d_{a}\right) ^{2}+f_{sa}^{\dagger }f_{sa}\right] ,
\end{equation}
with \( d_{a} \) given by Eq. (\ref{2.1}) and 
\begin{eqnarray}
f_{1} & \equiv  & \frac{1}{2}\left( e\left[ \phi _{3},\phi _{1}\right] -\mu \phi _{1}\right) ,\nonumber \\
f_{2} & \equiv  & \frac{1}{2}\left( e\left[ \phi _{3},\phi _{2}\right] +\mu \phi _{2}\right) ,\label{2.6} \\
f_{3} & \equiv  & \frac{1}{2}\left( e\left[ \phi _{1},\phi _{2}\right] -\mu _{3}\phi _{3}\right) .\nonumber 
\end{eqnarray}
This potential fulfills condition (\ref{2.2}). For this potential, the BPS
condition (\ref{2.4c}) is equivalent to the condition
\[
f_{s}=0,\, \, \, s=1,\, 2,\, 3\, .\]
 This is the potential of the bosonic part of \( \mathcal{N}=4 \) SYM theory
with some mass term deformations which break completely the supersymmetry. If
we set \( m=0 \), \( \mathcal{N}=1 \) supersymmetry is restored and we obtain
the potential considered in \cite{witten}. If further \( \mu _{3}=0 \), we
recover the potential of \( \mathcal{N}=2 \) with a massive hypermultiplet
in the adjoint representation. Finally, if also \( \mu =0 \), we obtain \( \mathcal{N}=4 \).
As usual, we shall denote by \( \mathcal{N}=2^{*} \), \( \mathcal{N}=1^{*} \),
and \( \mathcal{N}=0^{*} \) to the theories which are obtained by adding deformation
mass terms to \( \mathcal{N}=4 \) SYM theory.

Note that the term \( X_{a} \) is necessary if one wants to have a BPS string
which is not tensionless. Therefore this term generalizes the r\^ ole of Fayet-Iliopolous
terms in theories with \( U(1) \) factors by giving tension to the BPS string.
However, \( X_{a} \) in general breaks supersymmetry.

\section{Phases of the theory}

The vacua of the theory are solutions of
\begin{equation}
\label{3.1}
G_{\mu \nu }=D_{\mu }\phi _{s}=V=0\, .
\end{equation}
The condition \( V(\phi _{s})=0 \) is equivalent to 
\begin{equation}
\label{3.2}
d_{a}=0=f_{sa}\, .
\end{equation}
We shall only consider the theory in the weak coupling regime, and therefore
we shall not consider the quantum corrections to the potential. We are looking
for vacuum solutions which produce the symmetry breaking 
\[
G\, \rightarrow \, U(1)^{r}\, \rightarrow \, C_{G,}\]
where \( r \) is the rank of \( G \) and \( C_{G} \) its center. For the
first phase transition magnetic monopoles will appear. Then, in the second phase
transition magnetic flux tubes or strings (if \( C_{G} \) is non-trivial) will
appear and the monopoles will become confined. In order to produce this symmetry
breaking we shall look for vacuum solutions of the form 
\begin{eqnarray}
\phi _{1}^{\textrm{vac}} & = & a_{1}T_{+}\, ,\nonumber \\
\phi _{2}^{\textrm{vac}} & = & a_{2}T_{-}\, ,\label{3.3} \\
\phi _{3}^{\textrm{vac}} & = & a_{3}T_{3}\, ,\nonumber \\
W^{\textrm{vac}}_{\mu } & = & 0\, ,\nonumber 
\end{eqnarray}
where \( a_{1} \) and \( a_{2} \) are complex constants, \( a_{3} \) is a
real constant, and 
\begin{eqnarray*}
T_{3} & = & \delta \cdot H\, \, \, \, ,\, \, \, \, \, \delta \equiv \sum _{i=1}^{r}\frac{2\lambda _{i}}{\alpha _{i}^{2}}=\frac{1}{2}\sum _{\alpha >0}\frac{2\alpha }{\alpha ^{2}},\\
T_{\pm } & = & \sum _{i=1}^{r}\sqrt{c_{i}}E_{\pm \alpha _{i}},
\end{eqnarray*}
with \( \alpha _{i} \) and \( \lambda _{i} \) being simple roots and fundamental
weights, respectively, and 
\[
c_{i}\equiv \sum _{j=1}^{r}\left( K^{-1}\right) _{ij},\]
with \( K_{ij}=2\alpha _{i}\cdot \alpha _{j}/\alpha _{j}^{2} \) being the Cartan
matrix. The generators \( T_{\pm } \), \( T_{3} \) form the so called principal
\( SU(2) \) subalgebra of \( G \). The vacuum configuration \( \phi _{3}^{\textrm{vac}} \)
breaks \( G \) into \( U(1)^{\textrm{r}} \) and then \( \phi _{1}^{\textrm{vac}} \)
or \( \phi _{2}^{\textrm{vac}} \) breaks it further to \( C_{G} \). Note that
this is not the only possible vacuum configuration which produce the above symmetry
breaking. However, in this paper we shall restrict ourselves to analyzing this
configuration. We shall adopt the conventions
\begin{eqnarray*}
\left[ H_{i},E_{\alpha }\right]  & = & \alpha ^{i}E_{\alpha },\\
\left[ E_{\alpha ,}E_{-\alpha }\right]  & = & \frac{2\alpha }{\alpha ^{2}}\cdot H,
\end{eqnarray*}
where the upper index in \( \alpha ^{i} \) means the \( i \) component of
the root \( \alpha  \). Let 
\[
\alpha _{i}^{\textrm{v}}\equiv \frac{2\alpha _{i}}{\alpha ^{2}_{i}}\, \, \, ,\, \, \, \, \, \lambda _{i}^{\textrm{v}}\equiv \frac{2\lambda _{i}}{\alpha _{i}^{2}},\]
be the simple coroots\footnote{
In this paper the definition of \( \alpha _{i}^{\textrm{v}} \) differs from
the one adopted in \cite{kb} and \cite{k} by a factor of 2.
} and fundamental coweights, respectively. Then using the relations
\begin{eqnarray*}
\lambda _{j}^{\textrm{v}} & = & \alpha ^{\textrm{v}}_{i}\left( K^{-1}\right) _{ij},\\
\lambda _{i}^{\textrm{v}}\cdot \alpha _{j} & = & \delta _{ij},
\end{eqnarray*}
we obtain from the vacuum equations \( d=0=f_{s} \), that
\begin{eqnarray*}
\left( a_{3}-\frac{\mu }{e}\right) a_{i} & = & 0\, \, ,\, \, \, \, \textrm{for }i=1,2\, ,\\
a_{1}a_{2} & = & \frac{\mu _{3}a_{3}}{e}\, ,\\
ma_{3} & = & \left| a_{2}\right| ^{2}-\left| a_{1}\right| ^{2}\, .
\end{eqnarray*}

Independently of the values of the mass parameters, this system always has the
trivial solution \( a_{1}=a_{2}=a_{3}=0 \), which corresponds to the vacuum
in which the \( G \) is unbroken. Let us analyze other possible vacuum solutions
in which \( G \) is broken.

From this system we can conclude that if \( \mu =0 \), there exist nonvanishing
solution only if \( \mu _{3}=0=m \), which means that we recover \( \mathcal{N}=4 \)
SYM theory. In this case, \( a_{1}=0=a_{2} \) and \( a_{3} \) can be arbitrary
which implies that \( G \) is broken to \( U(1)^{r} \) if \( a_{3} \) is
nonvanishing. But then if we add an \( \mathcal{N}=1 \) or \( \mathcal{N}=0 \)
deformation to the \( \mathcal{N}=4 \) potential, by considering either \( \mu _{3} \)
or \( m \) nonvanishing, then the only solution is the trivial \( a_{1}=a_{2}=a_{3}=0 \)
and \( G \) returns to be unbroken. Therefore monopole confinement does not
happen when \( \mu =0 \), at least for vacuum configurations such as Eq. (\ref{3.3}). 

For the \( \mathcal{N}=2^{*} \) theory, in which \( \mu \neq 0 \) and \( \mu _{3}=0=m \),
the situation is as in the \( \mathcal{N}=4 \) case, with the solution \( a_{1}=0=a_{2} \)
and \( a_{3} \) arbitrary, which results in \( G \) broken to \( U(1)^{r} \)
for \( a_{3}\neq 0 \). Let us analyze the vacuum solutions when we add deformation
terms to \( \mathcal{N}=2^{*} \)

\subsubsection*{i) Adding the \protect\( \mathcal{N}=1\protect \) deformation term (\protect\( \mu \neq 0\protect \),
\protect\( \mu _{3}\neq 0\protect \), and \protect\( m=0\protect \)).}

In this case, there are non trivial solutions satisfying
\begin{eqnarray*}
a_{3} & = & \frac{\mu }{e},\\
a_{1}a_{2} & = & \frac{\mu _{3}\mu }{e},\\
|a_{1}|^{2} & = & |a_{2}|^{2},
\end{eqnarray*}
 which results in a vacuum which breaks \( G\, \rightarrow \, C_{G} \).

\subsubsection*{ii) Adding the \protect\( \mathcal{N}=0\protect \) deformation term (\protect\( \mu \neq 0\protect \),
\protect\( \mu _{3}=0\protect \), and \protect\( m\neq 0\protect \) ). }

If \( \mu _{3}=0 \), then either \( a_{1}=0 \) or \( a_{2}=0 \). We shall
take \( a_{1}=0 \). Then, there are two possible situations:

\begin{itemize}
\item \( m\mu <0 \) \( \Rightarrow  \) If we consider \( a_{2}\neq 0 \), then \( a_{3}=\mu /e \)
which would imply \( |a_{2}|^{2}<0 \). Therefore in this case, we must have
\( a_{2}=0=a_{3} \) and \( G \) remains unbroken.
\item \( m\mu >0 \) \( \Rightarrow  \) In this case there is the nontrivial solution

\begin{eqnarray}
a_{3} & = & \frac{\mu }{e}\, \, ,\\
\left| a_{2}\right| ^{2} & = & \frac{m\mu }{e}\, \, ,\label{3.7} 
\end{eqnarray}
which also results in a vacuum which breaks \( G\, \rightarrow \, C_{G} \). 
\end{itemize}

\subsubsection*{iii) Adding the \protect\( \mathcal{N}=1\protect \) and \emph{\protect\( \mathcal{N}=0\protect \)}
deformation terms (\protect\( \mu \neq 0\protect \), \protect\( \mu _{3}\neq 0\neq m\protect \)).}

In this case there are nontrivial solutions with
\[
a_{3}=\frac{\mu }{e}\]
and \( a_{1} \) and \( a_{2} \) satisfying 
\begin{eqnarray*}
a_{1}a_{2} & = & \frac{\mu _{3}\mu }{e}\, ,\\
\frac{m\mu }{e} & = & |a_{2}|^{2}-|a_{1}|^{2}\, .
\end{eqnarray*}
Once more \( G \) is broken to \( C_{G} \).

In summary, in the \( \mathcal{N}=4 \) and \( \mathcal{N}=2^{*} \) theory,
the gauge group \( G \) can be broken to \( U(1)^{r} \) which corresponds
to the Coulomb phase. If we add to \( \mathcal{N}=2^{*} \) a \( \mathcal{N}=1 \)
or \( \mathcal{N}=0 \) deformation (or both), the gauge group can be further
broken to \( C_{G} \), which gives rise to the Higgs or color superconducting
phase. Let us analyze each of these phases in the next sections.

\section{Coulomb phase}

In this phase \( G \) is broken to \( U(1)^{r} \) and there exist solitonic
monopole solutions. As we have seen, this phase can only occur for the \( \mathcal{N}=4 \)
and \( \mathcal{N}=2^{*} \) cases. That could happen, for example, for energy
scales in which one can consider \( \mu _{3}=0=m \). In this phase \( a_{1}=0=a_{2} \)
and \( a_{3}\neq 0 \). In principle \( a_{3} \) is an arbitrary nonvanishing
constant. However, we shall fix 
\[
a_{3}=\frac{\mu }{e}\]
in order to have the same value as in the Higgs phase. The vacuum solution \( \phi _{3}^{\textrm{vac}} \)
singles out a particular \( U(1) \) direction which we call \( U(1)_{\delta } \).
Since for any root \( \alpha  \), \( \delta \cdot \alpha \neq 0 \), we can
construct a monopole solution for each root \( \alpha  \). The asymptotic field
configuration for the monopole associated to the root \( \alpha  \) can be
written as \cite{bais}
\begin{eqnarray}
\phi _{3}(\theta ,\varphi ) & = & g_{\alpha }(\theta ,\varphi )\phi _{3}^{\textrm{vac}}g_{\alpha }(\theta ,\varphi )^{-1}=a_{3}g_{\alpha }(\theta ,\varphi )\delta \cdot Hg_{\alpha }(\theta ,\varphi )^{-1}\, ,\nonumber \\
B_{i}(\theta ,\varphi ) & = & \frac{r_{i}}{er^{2}}g_{\alpha }(\theta ,\varphi )T_{3}^{\alpha }g_{\alpha }(\theta ,\varphi )^{-1}\, ,\label{4.1} \\
\phi _{1}(\theta ,\varphi ) & = & \phi _{2}(\theta ,\varphi )=0\, ,\nonumber 
\end{eqnarray}
where 
\[
g_{\alpha }(\theta ,\varphi )=\exp (i\varphi T_{3}^{\alpha })\, \exp (i\theta T_{2}^{\alpha })\, \exp (-i\varphi T_{3}^{\alpha })\, ,\]
with the \( SU(2) \) generators
\[
T^{\alpha }_{1}=\frac{E_{\alpha }+E_{\alpha }}{2}\, \, ,\, \, \, \, T_{2}^{\alpha }=\frac{E_{\alpha }-E_{-\alpha }}{2i}\, \, ,\, \, \, \, T_{3}^{\alpha }=\frac{1}{2}\alpha ^{\textrm{v}}\cdot H.\]
 The associated monopole magnetic charge is 
\begin{equation}
\label{4.2}
g\equiv \frac{1}{|\phi _{3}^{\textrm{vac}}|}\oint dS_{l}\textrm{Tr}\left[ \textrm{Re}\left( \phi _{3}\right) B_{l}\right] =\frac{2\pi }{e}\frac{\delta \cdot \alpha ^{\textrm{v}}}{|\delta |},
\end{equation}
with 
\[
\delta ^{2}=\frac{h^{\textrm{v}}\left( h^{\textrm{v}}+1\right) r}{24}\, ,\]
where \( h^{\textrm{v}} \) is the dual Coxeter number of \( G \). Clearly
\( g \) is equal to the monopole magnetic flux in the \( U(1)_{\delta } \)
direction, \( \Phi _{\textrm{mon}} \). It is also convenient to define magnetic
fluxes associated to each \( U(1) \) factor of the unbroken group \( U(1)^{r} \).
As explained in \cite{go review}, in the sphere with \( r\rightarrow \infty  \),
the little group of \( \phi _{3}(\theta ,\varphi ) \) varies within \( G \)
by conjugation with the gauge transformations \( g_{\alpha }(\theta ,\varphi ) \).
Therefore the magnetic flux associated with the Cartan generator \( \lambda ^{\textrm{v}}_{i}\cdot H \),
\( i=1,\, 2,\, ...,\, r \), can be defined as 
\begin{equation}
\label{4.3}
\Phi _{\textrm{mon}}^{(i)}\equiv \oint dS_{_{l}}\textrm{Tr}\left[ g_{\alpha }(\theta ,\varphi )\lambda ^{\textrm{v}}_{i}\cdot Hg_{\alpha }(\theta ,\varphi )^{-1}B_{l}\right] =\frac{2\pi }{e}\lambda ^{\textrm{v}}_{i}\cdot \alpha ^{\textrm{v}}\, ,
\end{equation}
which are topologically conserved charges \cite{eweinberg}.

These are BPS monopoles with masses given by the central charge of the \( \mathcal{N}=2 \)
algebra \cite{witten olive, sw2}. For monopoles with vanishing fermion number,
their masses are \( M_{\textrm{mon}}=|g||\phi ^{\textrm{vac}}_{3}| \). Not
all of these monopoles are stable. The stable or fundamental are the ones with
lightest masses \cite{eweinberg}. For the present symmetry breaking, the fundamental
monopoles are associated with the simple roots for the simple-laced algebras
or to the long simple roots for the non-simply-laced ones. Their masses are
\begin{equation}
\label{4.4}
M^{\textrm{L}}_{\textrm{mon}}=\frac{2\pi }{e|\delta |}|\phi _{3}^{\textrm{vac}}|.
\end{equation}
Note that, since \( G \) is completely broken to \( U(1)^{\textrm{r}} \),
differently from \cite{k}, here the stable monopoles do not fill representations
of a non-Abelian unbroken group.

\section{Higgs or color superconducting phase}

In the Higgs or color superconducting phase, \( G \) is broken to its center
\( C_{G} \). That can happen when \( \mathcal{N}=2^{*} \) is broken by a \( \mathcal{N}=1 \)
or \( \mathcal{N}=0 \) deformation term (or both). In this phase, the monopole
chromomagnetic flux lines cannot spread out radially over space. A phenomenon
like that is expected to happen in the interior of very dense neutron stars
\cite{colorsupercondunting}. However, since for simply connected \( G \)
\begin{equation}
\label{5.0}
\Pi _{1}\left( G/C_{G}\right) =C_{G}.
\end{equation}
 If \( C_{G}=Z_{N} \), these flux lines can form topologically nontrivial \( Z_{N} \)
strings. Then, the monopoles of \( \mathcal{N}=2^{*} \) become confined in
this phase, as we shall show below.

In order to have finite string tension, these string solution must satisfy the
vacuum equations asymptotically, which implies that 
\begin{eqnarray*}
\phi _{s}(\varphi ,\rho \rightarrow \infty ) & = & g(\varphi )\phi _{s}^{\textrm{vac}}g(\varphi )^{-1},\\
W_{I}(\varphi ,\rho \rightarrow \infty ) & = & g(\varphi )W_{I}^{\textrm{vac}}g(\varphi )^{-1}-\frac{1}{ie}\left( \partial _{I}g(\varphi )\right) g(\varphi )^{-1},
\end{eqnarray*}
 where \( \rho  \) is the radial coordinate and capital Latin letters \( I,J \)
denote the coordinates \( 1 \) and \( 2 \); \( \phi _{s}^{\textrm{vac}} \)
and \( W_{I}^{\textrm{vac}} \) are given by Eq. (\ref{3.3}) and \( g(\varphi )\in G \).
In order for the field configuration to be single valued, \( g(\varphi +2\pi )g(\varphi )^{-1}\in C_{G} \).
Considering 
\[
g(\varphi )=\exp i\varphi M\, ,\]
 then \( \exp 2\pi iM\in C_{G} \). That implies that \( M \) must be diagonalizable
and we shall consider that 
\[
M=\omega \cdot H.\]
Then, in order to \( \exp 2\pi i\omega \cdot H\in C_{G} \), 
\[
\omega =\sum _{i=1}^{r}l_{i}\lambda ^{\textrm{v}}_{i},\]
where \( l_{i} \) are integer numbers; that is, \( \omega  \) must be a vector
in the coweight lattice of \( G \), which has the fundamental coweights \( \lambda _{i}^{\textrm{v}} \)
as basis vectors, and is equivalent to the weight lattice \( \Lambda _{\textrm{w}}\left( \widetilde{G}^{\textrm{v}}\right)  \)
of the covering group of the dual group \( \widetilde{G}^{\textrm{v}} \) \cite{gno}.
In principle, we could have other possibilities for \( M \) like some combinations
of step operators, which however we shall not discuss here. 

With the above choice for \( g(\varphi ) \), the asymptotic string configuration
can be written as 
\begin{eqnarray}
\phi _{s}(\varphi ,\rho \rightarrow \infty ) & = & e^{i\varphi \omega \cdot H}\phi _{s}^{\textrm{vac}}e^{-i\varphi \omega \cdot H},\label{5.1} \\
W_{I}(\varphi ,\rho \rightarrow \infty ) & = & \frac{\varepsilon _{IJ}x^{J}}{e\rho ^{2}}\omega \cdot H\, \, \, ,\, \, \, \, I=1,2.\label{5.2} 
\end{eqnarray}
 Note that not all of these strings are necessarily stable.

We can consider the string ansatz
\begin{eqnarray*}
\phi _{1}\left( \varphi ,\rho \right)  & = & e^{i\varphi \omega \cdot H}\sum _{i=1}^{r}\left( f^{i}_{1}(\rho )E_{\alpha _{i}}\right) e^{-i\varphi \omega \cdot H}=\sum _{i=1}^{r}e^{i\varphi \omega \cdot \alpha _{i}}f^{i}_{1}(\rho )E_{\alpha _{i}},\\
\phi _{2}\left( \varphi ,\rho \right)  & = & e^{i\varphi \omega \cdot H}\sum _{i=1}^{r}\left( f^{i}_{2}(\rho )E_{-\alpha _{i}}\right) e^{-i\varphi \omega \cdot H}=\sum _{i=1}^{r}e^{-i\varphi \omega \cdot \alpha _{i}}f^{i}_{2}(\rho )E_{-\alpha _{i}},\\
\phi _{3}\left( \varphi ,\rho \right)  & = & e^{i\varphi \omega \cdot H}\sum _{i=1}^{r}\left( f^{i}_{3}(\rho )\lambda _{i}^{\textrm{v}}\cdot H\right) e^{-i\varphi \omega \cdot H}=\sum _{i=1}^{r}f_{3}^{i}(\rho )\lambda _{i}^{\textrm{v}}\cdot H,\\
W_{I}\left( \varphi ,\rho \right)  & = & \frac{\varepsilon _{IJ}x^{I}}{e}\sum _{i=1}^{r}g_{i}(\rho )\lambda _{i}^{\textrm{v}}\cdot H,\\
W_{0}\left( \varphi ,\rho \right)  & = & 0=W_{3}\left( \varphi ,\rho \right) ,
\end{eqnarray*}
which results that 
\[
B_{3}\left( \varphi ,\rho \right) =\frac{1}{e\rho ^{2}}\sum _{i=1}^{r}\lambda ^{\textrm{v}}_{i}\cdot H\frac{\partial g_{i}\left( \rho \right) }{\partial \rho }.\]
These functions must satisfy the boundary condition
\begin{eqnarray*}
f^{i}_{n}(\rho \rightarrow \infty ) & = & a_{n}\sqrt{c_{i}}\, \, ,\, \, \, \textrm{for }n=1,2\, \, ,\\
f^{i}_{3}(\rho \rightarrow \infty ) & = & a_{3\, \, ,}\\
g_{i}(\rho \rightarrow \infty ) & = & l_{i}\, \, ,
\end{eqnarray*}
for \( i=1,\, ...,\, r \), in order to recover the asymptotic configuration
(\ref{5.1}), (\ref{5.2}) and 
\begin{eqnarray*}
f_{n}^{i}(\rho =0) & = & 0,\, \, \, \, \, \textrm{for }n=1,2\textrm{ and }i\textrm{ such that }\omega \cdot \alpha _{i}=l_{i}\neq 0\, \, \, ,\\
g_{i}(\rho =0) & = & 0,\, \, \, \, \, \textrm{for }i=1,\, ...,\, r\, \, \, ,
\end{eqnarray*}
in order to eliminate the singularities at \( \rho =0 \). One can put the ansatz
in the BPS conditions (\ref{2.4a})-(\ref{2.4d}) and obtain first order differential
equations similar to the ones in \cite{kb}. Otherwise one can put in the equations
of motion for the non-BPS cases. As in \cite{kb}, the BPS conditions are consistent
with the equations of motion only when \( m \) vanishes. However, in the case
in which \( \mathcal{N}=2^{*} \) is broken by an \( \mathcal{N}=0 \) deformation,
we must take the limit \( m\rightarrow 0 \) in order to maintain the symmetry
breaking \( G\, \rightarrow \, C_{G} \), similarly to the Prassard-Sommerfeld
limit \cite{prasar sommerfield} for the BPS monopole. 

As in \cite{znvortices} we shall consider that \( \phi _{3} \) is constant
and equal to its asymptotic value; i. e.,
\begin{equation}
\label{5.3}
\phi _{3}\left( \varphi ,\rho \right) =\frac{\mu }{e}T_{3}.
\end{equation}
 For the BPS string, using the above ansatz and boundary conditions, one obtains
Eq. (\ref{5.3}) directly from the BPS condition \( D_{\mp }\phi _{3}=0 \),
which implies that \( f_{3}^{i}(\rho )=\textrm{const}=a_{3}=\mu /e \). Hence
the lower bound for the string tension given by Eq. (\ref{2.3}), for \( X=em\, \textrm{Re}(\phi _{3})/2 \),
can be written as
\begin{equation}
\label{5.4}
T\geq \frac{me}{2}\left| \phi ^{\textrm{vac}}_{3}\right| \left| \Phi _{\textrm{st}}\right| =\frac{m\mu }{2}\left| \delta \right| \left| \Phi _{\textrm{st}}\right| 
\end{equation}
 where

\begin{equation}
\label{5.5}
\Phi _{\textrm{st}}\equiv \frac{1}{\left| \phi _{3}^{\textrm{vac}}\right| }\int d^{2}x\textrm{Tr}\left( \textrm{Re}\left( \phi _{3}\right) B_{3}\right) =\frac{-e}{\mu \left| \delta \right| }\oint dl_{I}\textrm{Tr}\left( \textrm{Re}\left( \phi _{3}\right) W_{I}\right) =\frac{2\pi }{e}\frac{\delta \cdot \omega }{|\delta |}
\end{equation}
is the string flux in the \( U(1)_{\delta } \) direction. The bound in Eq.
(\ref{5.4}) holds for the BPS strings. For the case of \( \mathcal{N}=2^{*} \)
broken by an \emph{\( \mathcal{N}=0 \)} deformation, the limit \( m\, \rightarrow \, 0 \)
would imply \( T\, \rightarrow \, 0 \). Then, if one wants to have a BPS string
with finite \( T \), one should also take \( \mu \rightarrow \infty  \), similarly
to the case of the BPS \( Z_{k} \) strings in \cite{kb}. A similar limit was
considered in \cite{hannay strassler zaffaroni, vainshtein yung}. That is exactly
like the London limit in the Abelian Higgs theory describing superconductors
where one takes to infinity the mass of the scalar. On the other hand, when
\( \mathcal{N}=2^{*} \) is broken by an \( \mathcal{N}=1 \) deformation (i.e.
\( m=0 \), \( \mu \neq 0 \), and \( \mu _{3}\neq 0 \)), from Eq. (\ref{5.4})
we see that the BPS string will be tensionless. The same happens in general
for the BPS strings associated with a coweight \( \omega  \) such that \( \delta \cdot \omega =0 \),
and therefore \( \Phi _{\textrm{st}}=0 \). 

Similarly to the monopole, we can define string fluxes associated with each
Cartan element \( \lambda ^{\textrm{v}}_{i}\cdot H \):
\begin{equation}
\label{5.6}
\Phi ^{(i)}_{\textrm{st}}\equiv \int d^{2}x\textrm{Tr}\left[ \lambda _{i}^{\textrm{v}}\cdot H\, B_{3}\right] =\frac{2\pi }{e}\lambda ^{\textrm{v}}_{i}\cdot \omega \, .
\end{equation}
Therefore, from Eq. (\ref{5.5}) or (\ref{5.6}), we can conclude that the string
fluxes take values in the coweight lattice of \( G \). Let us now check if
the magnetic fluxes of the monopoles are compatible with the ones of the strings.
Since an arbitrary coroot \( \alpha ^{\textrm{v}} \) can always be expanded
in the coweight basis as \( \alpha ^{\textrm{v}}=\sum _{i=1}^{r}n_{i}\lambda _{i}^{\textrm{v}} \)
where \( n_{i} \) are integer numbers, one can conclude that the magnetic fluxes
(\ref{4.2}) or (\ref{4.3}) of the monopoles can in principle be expressed
as an integer linear combination of the string fluxes (\ref{5.5}) or (\ref{5.6}).
Therefore, in the Higgs phase, the monopole magnetic flux lines can no longer
spread radially over the space, since \( G \) is broken to the discrete group
\( C_{G} \). However, they can form one or more flux tubes or strings, and
the monopoles can become confined. We shall call this set of strings attached
to a monopole as confining strings. This set of confining strings must have
total flux given by Eq. (\ref{5.5}) or (\ref{5.6}) with \( \omega =\alpha ^{\textrm{v}} \).
That means that this set of confining magnetic strings belongs to the trivial
topological sector of \( \Pi _{1}(G/C_{G}) \) since \( \exp 2\pi i\alpha ^{\textrm{v}}\cdot H=1 \)
in \( G \). The fact that the set of confining strings must belong to the trivial
sector is consistent with the fact that the set is not topologically stable
and therefore can terminate at some point. However, since it has a nonvanishing
flux, it must terminate in a magnetic source; i. e., a monopole. It is important
to stress the fact that a string configuration belonging to the topological
trivial sector does not imply that its flux must vanish as we can see from Eq.
(\ref{5.5}). All these results are generalizations of some well-known results
for the \( Z_{2} \) string of \( SU(2) \) Yang-Mills-Higgs theory, as explained
in \cite{hk review, Vilenkin}. In this theory there are at least two complex
scalars in the adjoint representation which produce the symmetry breaking \( SU(2)\, \rightarrow \, U(1)\, \rightarrow \, Z_{2} \),
similarly to our case. In the Higgs phase, string configurations can in principle
exist with flux \( 2\pi n/e \) for any integer \( n \), although only the
ones with \( n=\pm 1 \) are topologically stable. The ones with odd \( n \)
belong to the topologically nontrivial sector while the ones with even \( n \)
belong to the trivial sector. Therefore string configurations belonging to the
same topological sector do not have necessarily the same flux and therefore
are not related by (nonsingular) gauge transformations \cite{hk review, hk}.
The string configuration with \( n=2 \), belonging to the trivial sector and
which can be formed by two strings with \( n=1 \), is the one which can terminate
in the 't Hooft-Polyakov monopole with magnetic charge \( g=4\pi /e \) and
can produce the monopole-antimonopole confinement \cite{thooftreview}. In more
algebraic terms one can say that this set of integer numbers \( n \) forms
the coweight lattice \( \Lambda _{\textrm{w}} \) of \( SU(2) \), the subset
of even numbers \( 2n \) form the \( SU(2) \) coroot lattice \( \Lambda _{\textrm{r}} \),
and the quotient \( \Lambda _{\textrm{w}}/\Lambda _{\textrm{r}}\simeq Z_{2} \)
corresponds to the center of \( SU(2) \). Therefore this quotient has two elements
which are represented by the cosets \( \Lambda _{\textrm{r}} \) and \( 1+\Lambda _{\textrm{r}} \).
Each coset corresponds to a string topological sector, with \( \Lambda _{\textrm{r}} \)
been the trivial one. These results also holds for an arbitrary group \( G \),
where \cite{gno}
\begin{equation}
\label{5.7}
C_{G}\simeq \frac{\Lambda _{\textrm{w}}\left( \widetilde{G}^{\textrm{v}}\right) }{\Lambda _{\textrm{r}}\left( G^{\textrm{v}}\right) },
\end{equation}
with \( \Lambda _{\textrm{r}}(G^{\textrm{v}}) \) been the root lattice of the
dual group \( G^{\textrm{v}} \) or, equivalently, the coroot lattice of \( G \),
which has the simple coroots \( \alpha _{i}^{\textrm{v}} \) as basis vectors.
If the order of \( C_{G} \) is \( N \), then this quotient can be represented
by the \( N \) cosets 
\begin{equation}
\label{6.2}
\Lambda _{\textrm{r}}(G^{\textrm{v}})\, \, \textrm{and}\, \, \lambda _{i,\textrm{min}}^{\textrm{v}}+\Lambda _{\textrm{r}}(G^{\textrm{v}}),
\end{equation}
 where \( \lambda _{i,\textrm{min}}^{\textrm{v}} \) are the minimal fundamental
coweights of \( G \). A fundamental coweight is minimal if
\[
\lambda _{i,\textrm{min}}^{\textrm{v}}\cdot \psi =1,\]
where \( \psi  \) is the highest root and there exist exactly \( (N-1) \)
of them. The minimal coweights \( \lambda ^{\textrm{v}}_{i,\textrm{min}} \)
are associated with a special outer automorphism of the extended Dynkin diagrams
\cite{oliveturok1983}. For \( SU(N) \), all fundamental weights \( \lambda _{i} \)
are minimal. 

From Eqs. (\ref{5.0}) and (\ref{5.7}) it implies that 
\begin{equation}
\label{5.8}
\Pi _{1}(G/C_{G})=\frac{\Lambda _{\textrm{w}}\left( \widetilde{G}^{\textrm{v}}\right) }{\Lambda _{\textrm{r}}\left( G^{\textrm{v}}\right) },
\end{equation}
and we can conclude that each string topological sectors is associated with
a coset in Eq. (\ref{6.2}), with \( \Lambda _{\textrm{r}}(G^{\textrm{v}}) \)
being the trivial topological sector. It is important to note that for \( G=SU(N) \),
this result is equivalent to considering the string topological sectors to be
associated with the \( N \)-ality of the representations. However the above
result holds for arbitrary \( G \). 

Since the confining string configuration linking a monopole to an antimonopole
belongs to the trivial topological sector, it can break when it has enough energy
to create a new monopole-antimonopole pair. As was done in \cite{k}, one can
obtain a bound for the threshold length \( d^{\textrm{th}} \)for the string
breaking, using the relation
\begin{equation}
\label{5.9}
2M^{\textrm{L}}_{\textrm{mon}}=E^{\textrm{th}}=Td^{\textrm{th}}\geq \frac{me}{2}\left| \phi ^{\textrm{vac}}_{3}\right| \left| \Phi _{\textrm{st}}\right| d^{\textrm{th}}\, ,
\end{equation}
where \( E^{\textrm{th}} \) is the string threshold energy and \( M^{\textrm{L}}_{\textrm{mon}} \)
is the mass of the lightest monopoles, given by Eq. (\ref{4.4}). In the above
relation we used the string bound given by Eq. (\ref{5.4}) and did not consider
a possible energy term proportional to the inverse of the monopole distance,
known as the Lucher term. The modulus of the string flux, \( |\Phi _{\textrm{st}}| \),
must be equal to the modulus of the magnetic charges \( |g| \) of each confined
monopoles. Let us consider that \( |g|=2\pi |\delta \cdot \beta ^{\textrm{v}}|/|\delta | \)
with \( \beta ^{\textrm{v}} \) being an arbitrary coroot. Therefore one can
conclude from Eq. (\ref{5.9}), using Eq. (\ref{4.4}), that
\[
d^{\textrm{th}}\leq \frac{4}{me|\delta \cdot \beta ^{\textrm{v}}|}.\]

\section{Monopole confinement for \protect\( SU(N)\protect \) broken to \protect\( Z_{N}\protect \)}

Let us consider \( G=SU(N) \). Since \( SU(N) \) is simply laced, we do not
need to distinguish between weights and coweights, roots and coroots. We have
seen that the magnetic lines of a given monopole with magnetic charge \( g=2\pi \delta \cdot \alpha ^{\textrm{v}}/|\delta | \)
can form a set of flux tubes or strings. However, there are countless different
string configurations with this magnetic flux. It is not clear at the moment
which could be the preferable one. The most ``economical'' sets would be the
ones formed by a strings and an antistring as we shall see below. 

For \( SU(3) \), the quotient (\ref{5.7}), which is equivalent to \( C_{SU(3)}=Z_{3} \),
possesses three elements which can be represented by the cosets \( \Lambda _{\textrm{r}}(SU(3)) \),
\( \lambda _{1}+\Lambda _{\textrm{r}}(SU(3)) \) and \( \lambda _{2}+\Lambda _{\textrm{r}}(SU(3)) \).
One can, for example, construct string solutions associated with each of the
three weights \( \lambda _{1},\, \lambda _{1}-\alpha _{1},\, \lambda _{1}-\alpha _{1}-\alpha _{2} \)
of the fundamental representation. Since all of them belong to the coset \( \lambda _{1}+\Lambda _{\textrm{r}} \),
these string solutions belong to the same topological sector. However, one can
observe from Eq. (\ref{5.5}) that they do not have the same flux \( \Phi _{\textrm{st}} \),
similarly to the \( Z_{2} \) strings of \( SU(2) \) theory. Therefore these
string solutions are \textit{not} related by gauge transformations since \( \Phi _{\textrm{st}} \)
is gauge invariant. One can construct the corresponding antistring solutions
associated with the negative of these weights, which form the complex-conjugated
representation \( \overline{3} \) and which belong to the coset \( \lambda _{2}+\Lambda _{\textrm{r}} \).
The magnetic fluxes of the monopoles associated with the six nonvanishing roots
of \( SU(3) \) can easily be written using these strings in the following way:
for the monopole \( \alpha _{1} \) we can attach the strings \( \lambda _{1} \)
and \( -\lambda _{1}+\alpha _{1} \). For the monopole \( \alpha _{2} \) we
can attach strings \( \lambda _{1}-\alpha _{1} \) and \( -\lambda _{1}+\alpha _{1}+\alpha _{2} \).
For the monopole \( \alpha _{1}+\alpha _{2} \) we can attach the strings \( \lambda _{1} \)
and \( -\lambda _{1}+\alpha _{1}+\alpha _{2} \). And similarly for the other
three monopoles associated with the negative roots, just changing the signs.
The remaining three combinations of strings and antistring have vanishing fluxes
\( \Phi ^{(i)}_{\textrm{st}} \). 

One could draw the above set of strings attached to monopoles as shown in Fig.1,
where the circles represent the monopoles and the arrows are the string flux
\( \Phi ^{(i)}_{\textrm{st}} \). We represented the strings associated with
weights in the fundamental representation by an arrow going out of the monopole
and for the antistrings we reversed the sense of the arrow and simultaneously
changed the sign of the weight. Then, in addition to the monopole-antimonopole
pairs one could also conjecture about the formation of a confined system with
the monopoles \( \alpha _{1} \), \( \alpha _{2} \) and \( -\alpha _{1}-\alpha _{2} \)
as shown in Fig. 2. Note that since these monopoles are not expected to fill
the three dimensional fundamental representation of \( SU(3) \), that system
is not exactly like a baryon. With this configuration of monopoles with strings
attached, one could also think of putting one string in the north pole and the
on the other in the south pole, forming a configuration similar to the bead
described in \cite{hk}.

In principle one could also think of attaching to the monopole \( \alpha _{1} \)
the strings \( 2\lambda _{1} \) and \( -2\lambda _{1}+\alpha _{1} \) which
belong to the six-dimensional symmetric tensor representation and its complex
conjugate. However, one can conclude directly from the expression for the fluxes
and string tension that, in the BPS case, the string \( 2\lambda _{1} \) can
decay in two strings \( \lambda _{1} \) and the string \( -2\lambda _{1}+\alpha _{1} \)
can decay in the strings \( -\lambda _{1} \) and \( -\lambda _{1}+\alpha _{1} \).
A similar result has been conjectured in some different theories \cite{strassler, konishi, shifman yung}.

One can easily extend the above construction of strings attached to monopoles
to the \( SU(N) \) case. In this case the quotient (\ref{5.7}) has \( N \)
elements which can be represented by the cosets 
\[
\Lambda _{\textrm{r}}(SU(N))\, \, \textrm{and}\, \, \lambda _{i}+\Lambda _{\textrm{r}}(SU(N)),\, \, i=1,2,\, ...\, ,\, N-1.\]
 The representation corresponding to the fundamental weight \( \lambda _{k} \)
can be obtained by the antisymmetric tensor product of \( k \) fundamental
representations associated with \( \lambda _{1} \). Once more one can consider
the strings associated with the weights of the fundamental \( N \) dimensional
representation, 
\[
\lambda _{1}\, \, \, \, \textrm{and}\, \, \, \, \lambda _{1}-\sum _{j=1}^{l}\alpha _{j}\, \, \, ,\, \, \, \, l=1,\, 2,\, ...,\, N-1,\]
which belong to the \( \lambda _{1}+\Lambda _{\textrm{r}}(SU(N)) \) coset,
and with the negative of these weights which form the complex-conjugated representation
\( \overline{N} \), which belongs to the \( \lambda _{N-1}+\Lambda _{\textrm{r}}(SU(N)) \)
coset. Since \( N\times \overline{N}=\textrm{adj}+1 \), one can write the fluxes
of monopoles (not all of them stable) associated with the \( N(N-1) \) nonvanishing
roots, in terms of a string and an antistring. For example for the monopole
associated with the \( SU(N) \) root \( \alpha _{p}+\alpha _{p+1}+\cdots +\alpha _{p+q} \)
one could attach strings associated with the weights \( \lambda _{1}-\alpha _{1}-\alpha _{2}-\cdots -\alpha _{p-1} \)
and \( -\lambda _{1}+\alpha _{1}+\cdots +\alpha _{p+q} \). As for the \( SU(3) \)
case, one could in principle form a confining configuration formed by the monopoles
associated with the \( N-1 \) simple roots \( \alpha _{i} \) and the negative
of the highest root \( -\alpha _{1}-\alpha _{2}-\cdots -\alpha _{N-1} \).

Since, for \( SU(N) \), \( \lambda _{N-k}=-\lambda _{k}+\beta  \), where \( \beta \in \Lambda _{\textrm{r}}(SU(N)) \),
each weight in the coset \( \lambda _{N-k}+\Lambda _{\textrm{r}}(SU(N)) \)
is the negative of a weight in \( \lambda _{k}+\Lambda _{\textrm{r}}(SU(N)) \),
and therefore the bound of a string tension, given by Eq. (\ref{5.4}), associated
with a weight in \( \lambda _{k}+\Lambda _{\textrm{r}}(SU(N)) \) should be
equal to the one associated with the negative weight in \( \lambda _{N-k}+\Lambda _{\textrm{r}}(SU(N)) \).

\section{String tension and Casimir scaling law}

The string tension is one of the main quantities to be determined in quark confinement
in QCD. In these last 20 years quite a lot of work has been done trying to determine
this quantity. There are mainly two conjectures for the string tension: the
``Casimir scaling law'' \cite{casimir} and the ``sine law'' \cite{douglas shenker}.
In these two conjectures the gauge group \( G=SU(N) \) is considered and a
string in the representation associated with the fundamental weight \( \lambda _{k} \)
which can be obtained by the antisymmetric tensor product of \( k \) fundamental
representations associated with \( \lambda _{1} \).  For the Casimir scaling
conjecture, the string tension should satisfy 
\begin{equation}
\label{7.1}
T_{k}=T_{1}\frac{k(N-k)}{N-1},\, \, \, \, \, \, k=1,\, 2,\, ...,\, N-1,
\end{equation}
 where \( T_{1} \) would be the string tension in the \( \lambda _{1} \) fundamental
representation. On the other hand, for the sine law conjecture, 
\[
T_{k}=T_{1}\frac{\sin \left( \frac{\pi k}{N}\right) }{\sin \left( \frac{\pi }{N}\right) },\, \, \, \, \, \, k=1,\, 2,\, ...,\, N-1.\]
 There are some papers \cite{hannay strassler zaffaroni, strassler, shifman, sine}
using different approaches like MQCD, AdS/CFT, etc., giving some support to
this last conjecture. On the other hand, several lattice studies \cite{lattice}
have appeared in the literature in the last years giving support to both conjectures. 

All these conjectures are concerned with the chromoelectric strings. However,
as we mentioned in the introduction, one expects that chromomagnetic strings
could be related to chromoelectric strings by a duality transformation. Therefore
one could ask if the tensions of our chromomagnetic strings satisfy one of the
two conjectures. 

Let us start with a general gauge group \( G \). From Eqs. (\ref{5.4}) and
(\ref{5.5}), we obtain that the string tension satisfies the bound
\[
T_{\omega }\geq \frac{m\mu \pi }{e}\left| \delta \cdot \omega \right| ,\]
where \( \omega  \) must belong to one of the cosets (\ref{6.2}). Let 
\[
\omega =\lambda _{k}^{\textrm{v}}-\beta _{\omega },\]
where \( \lambda _{k}^{\textrm{v}} \) is a minimal fundamental coweight and
\( \beta _{\omega }\in \Lambda _{\textrm{r}}(G^{\textrm{v}}) \). Remembering
that \( \lambda _{k}^{\textrm{v}} \) is a fundamental weight of \( \widetilde{G}^{\textrm{v}} \)
and the quadratic Casimir associated with this fundamental representation is

\[
C(\lambda _{k}^{\textrm{v}})=\lambda _{k}^{\textrm{v}}\cdot \left( \lambda _{k}^{\textrm{v}}+2\delta \right) ,\]
it follows that
\begin{equation}
\label{7.2}
T_{\omega }\geq \frac{m\mu \pi }{e}\left| \frac{1}{2}\left[ C(\lambda _{k}^{\textrm{v}})-\lambda _{k}^{\textrm{v}}\cdot \lambda _{k}^{\textrm{v}}\right] -\delta \cdot \beta _{\omega }\right| .
\end{equation}
 Clearly for a string configuration in the trivial topological sector, i.e.
\( \omega =-\beta _{\omega }\in \Lambda _{\textrm{r}}(G^{\textrm{v}}) \), the
above string tension bound does not have the first term. 

Let us now consider \( G=SU(N) \). The quadratic Casimir associated with the
representation with fundamental weight \( \lambda _{k} \) is
\[
C(\lambda _{k})=\frac{N^{2}-1}{2N}\left[ \frac{k\left( N-k\right) }{N-1}\right] .\]
Moreover,
\[
\lambda _{k}=e_{1}+e_{2}+\cdots +e_{k}-\frac{k}{N}\sum _{j=1}^{N}e_{j},\]
where \( e_{i}\cdot e_{j}=\delta _{i,j} \). Therefore
\[
\lambda _{k}\cdot \lambda _{k}=\frac{k\left( N-k\right) }{N}.\]
 Hence, for \( SU(N) \),
\begin{equation}
\label{7.3}
T_{\lambda _{k}-\beta _{\omega }}\geq \frac{m\mu \pi }{e}\left| \frac{1}{2}\left( \frac{\left( N-1\right) ^{2}}{2N}\frac{k\left( N-k\right) }{N-1}\right) -\delta \cdot \beta _{\omega }\right| 
\end{equation}
 Therefore the first term on the right-hand-side of this inequality or, equivalently,
the BPS string tension associated with \( \omega =\lambda _{k} \) can be written
as 
\begin{equation}
\label{7.4}
T^{{\scriptsize \textrm{ BPS}}}_{\lambda _{k}}=T^{{\scriptsize \textrm{ BPS}}}_{\lambda _{1}}\frac{k\left( N-k\right) }{N-1},\, \, k=1,\, 2,\, ...,\, N-1,
\end{equation}
where 
\[
T_{\lambda _{1}}^{{\scriptsize \textrm{ BPS}}}=\frac{m\mu \pi }{2e}\frac{\left( N-1\right) ^{2}}{2N}\]
 is the BPS string tension associated with \( \omega =\lambda _{1} \). Hence
we explicitly showed that the BPS string tensions associated with an arbitrary
\( SU(N) \) fundamental weight \( \lambda _{k} \) satisfy the Casimir scaling
conjecture, given by Eq. (\ref{7.1}). However, in the Casimir scaling law conjecture
(and also in the sine law conjecture), it is believed that the string tension
should be the same for all weights in a given topological sector. But from Eq.
(\ref{7.2}) or (\ref{7.3}), we can see that the first term depends only on
the coset or topological sector but the second term is proportional to \( \delta \cdot \beta _{\omega } \)
and therefore depends explicitly on which weight is being considered. As we
have seen, that result is exactly like the \( SU(2) \) case, where the strings
in a given topological sector (i.e., \( n \) even or odd) do not have same
magnetic flux and consequently string tension. On the other, hand only the strings
with \( n=\pm 1 \) are stable and satisfy Eq. (\ref{7.4}). As we have mentioned
before, not all of the strings associated with weights in a given coset are
expected to be stable. Therefore, it would be interesting to determine the stability
conditions for these string solutions, similarly to what was done for the BPS
monopoles \cite{eweinberg} and BPS \( U(1) \) string solutions \cite{eweibergvortex}.

\vskip 0.2 in \noindent \textbf{\large Acknowledgments} \textbf{}

\noindent

I would like to thank T. Hollowood, T. Kibble, and M. Strassler for discussions
and FAPERJ for financial support.

\newpage

\vspace{0.3cm}
{\centering \resizebox*{0.95\textwidth}{!}{\rotatebox{270}{\includegraphics{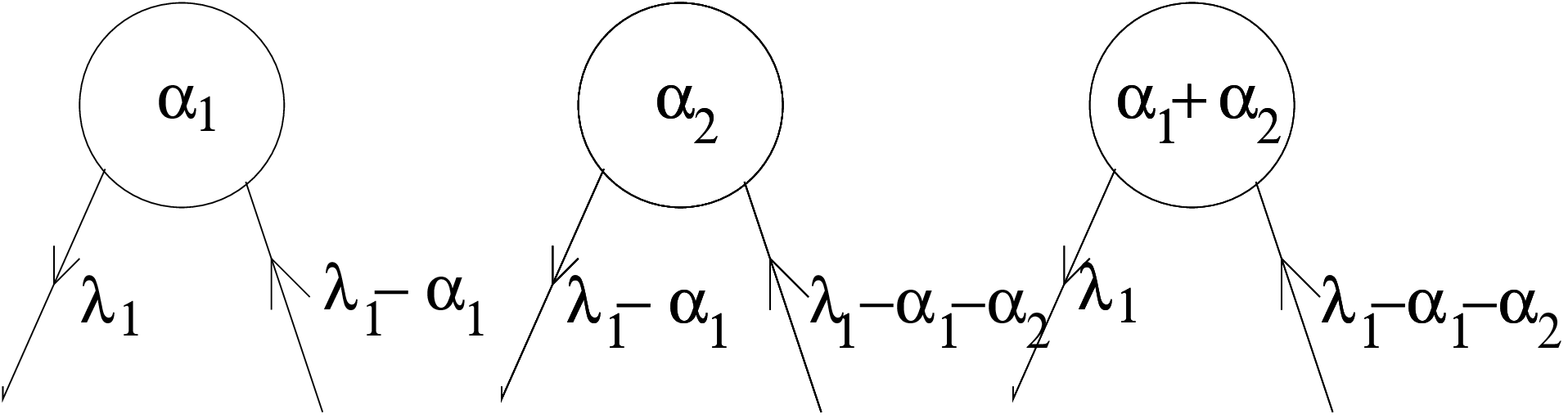}}} \par}

{\centering fig. 1\par}
\vspace{1cm}

\vspace{0.3cm}
{\centering \rotatebox{270}{\includegraphics{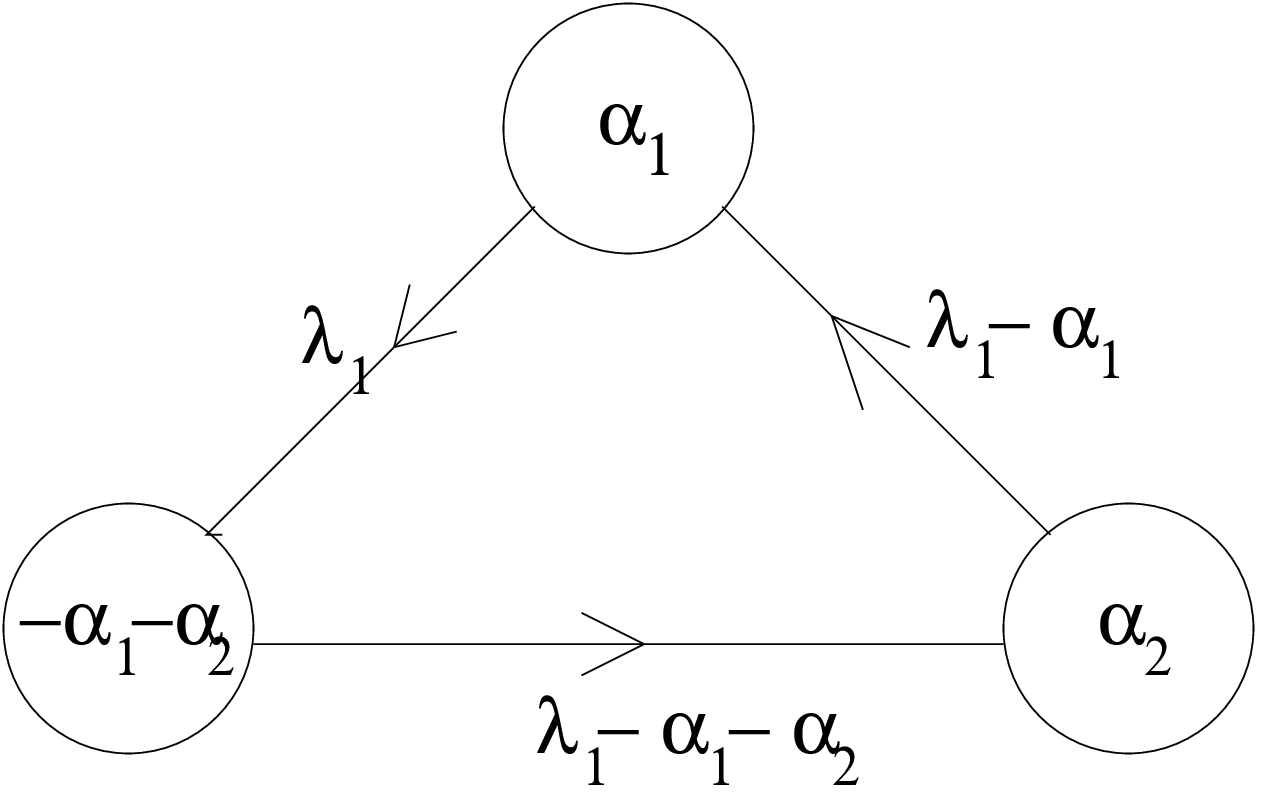}} \par}

{\centering fig. 2\par}
\vspace{0.3cm}

\end{document}